\documentclass[%
reprint,
groupedaddress,
 amsmath,amssymb,
 aps,
pra,
 ]{revtex4-2}
 
\usepackage{graphicx}
\usepackage{dcolumn}
\usepackage{bm}
\usepackage{lineno}

\def\bs#1{\bm{#1}}
\def\ve#1{\bs{\mathbf{#1}}}
\def\bal#1\eal{\begin{align}#1\end{align}}
\def\balx#1\ealx{\begin{align*}#1\end{align*}}
\def\be#1\ee{\begin{align}#1\end{align}}

\def\d{\mathrm{d}}
\def\rb{\ve{r}}

\def\mub{\bm{\mu}}
\def\nub{\bm{\nu}}
\def\etab{\ve{\eta}}

\def\xb{\ve{x}}

\usepackage{color}
\usepackage{soul}
\definecolor{violet}{rgb}{.56,0,1}
\definecolor{darkgreen}{rgb}{0,.6,0}

\begin{document}

\title{Super-resolution capacity of variance-based stochastic
  fluorescence microscopy }

\author{Simon Labouesse (1), Jérôme Idier (2), Marc Allain (3),
  Guillaume Giroussens (3), Thomas Mangeat (1) and Anne Sentenac (3)*}

\affiliation{(1) LITC Core Facility, Centre de Biologie Intégrative, Université de Toulouse, CNRS, UPS, 31062, Toulouse, France.}
\affiliation{(2) Nantes Université, École Centrale Nantes, LS2N, CNRS, Nantes, France.}
\affiliation{(3) Aix Marseille Univ, CNRS, Centrale Marseille, Institut Fresnel, Marseille, France.}


\begin{abstract}
  { \color{black} Improving the resolution of fluorescence microscopy beyond the diffraction
    limit can be achieved by acquiring and processing multiple images of the sample under different illumination conditions.
    One of the simplest techniques,
  Random Illumination Microscopy (RIM), forms
 the super-resolved image from the variance of 
  images obtained with random speckled illuminations. However, the
  validity of this process has not been fully theorized. In this work, we characterize mathematically the sample
  information contained in the variance of diffraction-limited speckled images as a function of the statistical
  properties of the illuminations. We show that an unambiguous two-fold resolution gain is obtained when the
  speckle correlation length coincides with the width of the observation point spread function.
  Last, we analyze the difference between the variance-based 
techniques using random speckled illuminations (as in RIM) and those obtained 
using random fluorophore activation (as in Super-resolution Optical Fluctuation 
Imaging, SOFI). }
\end{abstract}
\maketitle


{\color{black} The light intensity recorded by the camera of a fluorescence microscope
cannot exhibit spatial frequencies above
 $2/\lambda$ where $\lambda$
 is the wavelength of the emitted light.
 This low-pass filtering, due to the loss
 of the evanescent waves at the detector plane, cannot be
 circumvented.}
Therefore the challenge  of super-resolution 
imaging is to recover spatial frequencies of the sample fluorescence density
beyond $2/\lambda$ from data that are frequency limited to $2/\lambda$.
{\color{black} A widespread solution
consists in processing multiple images obtained by changing the illumination, like translating
focused spots \cite{Sheppard1990,Klar1999,Sheppard1988,Muller2010} or rotating and translating periodic light patterns
\cite{Heintzmann1999,Gustafsson2000}.}
The data processing of most techniques using structured illuminations requires the knowledge of the illumination patterns, either explicitely as in
Structured Illumination Microscopy \cite{Heintzmann1999,Gustafsson2000} or implicitely as in confocal or Image Scanning  microscopy \cite{Sheppard1990}.
In this context, Random Illumination Microscopy \cite{Mangeat2021,Dertinger2009,Ventalon2005} stands out as an exception
as it does not require the knowledge of the illumination patterns: the super-resolved image is formed from the variance of multiple
diffraction-limited images recorded under different random speckled illuminations.
While attractive because of its simplicity and
 significant image improvement \cite{Mangeat2021},
 RIM variance-based processing lacks a rigorous analysis of its resolution potential,
 the non-linearity of the variance being a significant
 obstacle to its derivation.
In this work, we study the sample information that can be extracted from 
 the variance of speckled images as a function of the statistical properties of the random illumination and
 we derive the condition under which the variance can provide a resolution gain.

{\color{black} To model the data provided by a fluorescence microscope under an inhomogeneous illumination, we introduce
  the point spread function of the microscope $h$ and
  the illumination intensity function $E$. Importantly, these two functions are defined
  at a macroscopic scale inside the sample,
  through the averaging over regions large enough to contain thousands of atoms
  (typically of the order of a thousand $\rm{nm}^3$), to wash out the microscopic fluctuations. 
In this context, we define the macroscopic fluorescence density $\rho$ such that
$V \rho(\rb)E(\rb)$ is the energy (detected by the camera)
of the fluorescent light emitted by a macroscopic volume $V$ centered about $\rb$.
Hereafter we neglect the Poisson noise.
The fluorescence density depends on the fluorophore concentration and the molecular brightness.}


With these definitions, the microscope image can be written as,
\bal
I(\rb)=\int \rho(\rb')E(\rb')h(\rb-\rb')\d\rb'
\label{model}
\eal
where  $\rb$ indicates a position in the
image domain that is conjugated to a point in the object domain.
This model can be applied to two- or three-dimensional imaging configurations.
In the Fourier space, Eq.~(\ref{model}) reads,
\bal
\tilde{I}(\nub)=\tilde{h}(\nub)
\int  \tilde{ \rho}  (\nub-\nub' )\tilde{E}(\nub')\d\nub'
\eal
where $\tilde{f}(\nub)=\int f(\rb) e^{-i2\pi\nub\cdot\rb} \rm{d}\rb$ stands for the Fourier transform of $f$.

If $E$ is a constant, as in a standard fluorescence microscope,  the recorded image
depends only on the sample spatial frequencies belonging to the support of the Optical Transfer Function (OTF)
$\tilde{h}$, noted $W_h$, which is
at best a disk of radius $2/\lambda$ (in the 2D case) or exhibits a torus-like shape in the 3D case \cite{sentenac2018}.

On the other hand, if $E$ is not a constant, and noting  $W_E$ the support of $\tilde{E}$, the recorded
image depends on the sample spatial frequencies in the
domain
\bal
W_{hE}= \{ \nub-\mub\,|\, \nub \in W_h, \mub \in W_E \},
  \label{base}
\eal
which is no more limited by $2/\lambda$ and corresponds to the
Fourier support of $hE$.

 Now, sensitivity to spatial frequencies of the sample outside $W_h$ is a necessary
but not a sufficient condition for being able to form a super-resolved image.
One also needs a technique for extracting the high
spatial frequencies of the sample from the diffraction-limited images $I_n$ obtained for various
illumination intensities $E_n$.
{\color{black} For example, in an ideal confocal microscope, the $n^{th}$ illumination is a focal spot centered about $\rb_n$ and
the measured data is $I_n(\rb_n)= \int \rho(\rb')E_n(\rb')h(\rb_n-\rb')\d\rb'$. If one neglects the Stokes shift and forms
the illumination through the same optical path as the collection, one gets (using the reciprocity theorem \cite{sentenac2018})
$E_n(\rb)=h(\rb_n-\rb)$ and
\be
I_n(\rb_n)= \int \rho(\rb')h^2(\rb_n-\rb')\d\rb'
\ee
which permits to recover the fluorescence density over $W_{hE}=W_{h^2}$ with
$W_{h^2}=\{ \nub-\nub'\,|\, \nub \in W_h, \nub' \in W_h\}$.
Note however that the super resolution capacity of an ideal confocal microscope requires
the use of an infinitely small pinhole which is impossible in practice. }
%
 Generally, when the different illumination intensities are known,
 the sample frequencies in $W_{hE}$ can be obtained explicitely
 from a linear combination of the diffraction-limited images (see \cite{Gustafsson2000} for exemple).
 
 On the other hand, when the illuminations are unknown speckles as in Random Illumination Microscopy,
 there is no simple way for recovering the sample frequencies from 
 the diffraction-limited images. 
 While a link between the (numerically intractable) full covariance
 of the images and the sample was derived in \cite{Idier2018}, there is currently no rigorous result concerning the
 sample information that can be extracted from the sole variance. 

 Let us consider an experiment in which multiple images of a sample are recorded under different speckled illuminations $E_n$ that are
 considered as different realizations of the same random process $E$.
  The variance of these diffraction-limited images,
  $V_{\rho} (\rb) = \mathbb{E}[I^2(\rb)]-\mathbb{E}[I(\rb)]^2$, where $\mathbb{E}$ stands for the ensemble average
  over the random illuminations, can be expressed as a function of the speckle autocovariance
$C(\rb,\rb')=\mathbb{E}[E(\rb) E(\rb')] - \mathbb{E}[E(\rb)]\mathbb{E}[E(\rb')$, as \cite{Idier2018},
\bal
\label{var}
V_{\rho}(\rb)=&\int\d\rb_1 \d\rb_2
\\
\nonumber
& h(\rb- \rb_1) \rho(\rb_1) C(\rb_1 ,\rb_2) \rho(\rb_2)h(\rb- \rb_2).
\eal

In the following, we assume that $E$ is
a second-order stationnary random process so that $C(\rb_1,\rb_2)=C(\rb_1-\rb_2)$ and
the Fourier support of $C$ is equal to the Fourier support of $E$, $W_E$\cite{priestley1981}.

Since the variance depends on the square
of raw images that are frequency limited to $W_h$,
its Fourier support corresponds to $W_{h^2}$.

On the other hand, each raw image 
is 
sensitive to the spatial frequencies of the
sample in $W_{hE}$,
so we expect the variance to be also sensitive to the
spatial frequencies of the sample in $W_{hE}$
(we show on a specific example in appendix~\ref{B}, that this is indeed the case).
These preliminary remarks bring out the
main question posed by variance-based super-resolved microscopy approaches: what sample information in $W_{hE}$
can we extract from an image that is frequency limited to $W_{h^2}$?
\\
We first consider RIM configuration in which the illumination
is performed through the same
  objective as the observation and the
  Stokes shift can be neglected. In this case, the speckle autocovariance $C$ is
  equal to the point spread function $h$ of the microscope
  \cite{Goodman2007,Mangeat2021} and $W_E$ is
  similar to $ W_h$.  Thus, the domain of sample spatial frequencies acting on
  RIM raw images, $W_{hE}$,  matches the Fourier support of the variance, $W_{h^2}$.
  In the following, we demonstrate that, indeed,
  if $h=C$, there is a bijection between the RIM variance and
 the sample frequencies within $W_{h^2}$.
 
 We start by noting that the Fourier transform of $C$, $\tilde{C}$, is positive as
 $C$ is an autocovariance function. We define $h_E$ such that
 $\tilde{h}_E=\sqrt{\tilde C}$. One easily shows that
 $ \int h_E(\rb_1 -\xb)h_E(\rb_2 -\xb)\d \xb= C(\rb_1-\rb_2)$.
 Using this decomposition of the speckle covariance,
 $V_{\rho}^{\textrm{RIM}}$ can be cast as, 
\bal
V_{\rho}^{\textrm{RIM}}(\rb)= B_{\rho,\rho}(\rb) 
\eal
where 
\bal
B_{U,V}(\rb) = \int M_U(\rb,\xb) M_V(\rb,\xb) \d\xb
\label{Buv}
\eal
with
\bal
M_V(\rb,\xb) = \int h(\rb- \rb_1) V(\rb_1) h_E( \rb_1 - \xb) \d\rb_1
\label{M}
\eal
and $U$ and $V$ are integrable real functions.
To pursue the demonstration, we need to point out several properties of $B_{U,V}$ and $M_V$.
First, it is easily seen that  $B_{U,V}$ is symmetric with respect to $(U,V)$, $B_{U,V}= B_{V,U}$
and bilinear, $B_{U+U',V+V'}= B_{U,V}+ B_{U',V'}+ B_{U,V'}+ B_{U',V}$.
Second, the Fourier transform of $M_V$ with respect to $(\rb,\xb)$,
\bal
\tilde{M}_{V}(\nub,\mub) = \tilde{h}(\nub)\tilde{V}(\mub+\nub) \sqrt{\tilde{C}(\mub)},
\label{FTM}
\eal
is bounded,  so $M_{V}$ is an analytic function.
Third, we show in appendix~\ref{A} that, if $h=C$,
\bal
  \int B_{U,V}(\rb) V(\rb) \d\rb = \int \left| M_{V}(\rb,\xb) \right|^2 U(\rb)\d\rb\d\xb.
  \label{norm}
\eal
\\
We now consider two fluorescence densities, $\rho_1(\rb) \geq 0$ and $\rho_2(\rb) \geq 0$
which have the same RIM variance, $B_{\rho_1,\rho_1}(\rb) = B_{\rho_2,\rho_2}(\rb)$.
Using the bilinearity and symetry of $B_{U,V}$, we can show that
$B_{\rho_1,\rho_1}-B_{\rho_2,\rho_2}=B_{\rho_1+\rho_2, \rho_1-\rho_2}=0$.
Thus, one obtains
\bal
    \label{eq_var_RIM}
    \int B_{\rho_1+\rho_2,\rho_1-\rho_2}(\rb) [ \rho_1(\rb) -\rho_2(\rb) ] \d\rb=0
\eal
    which, using Eq.~(\ref{norm}), implies,
\bal
\int \left| M_{\rho_1-\rho_2}(\rb,\xb) \right|^2 [ \rho_1(\rb) + \rho_2(\rb) ] \d\rb \d\xb=0.
\label{normbis}
\eal 
We now assume that $\rho_1 + \rho_2$ stays strictly positive in a non-empty open set $\Omega$.
In this case, Eq.~(\ref{normbis}) is satisfied if and only if $M_{\rho_1-\rho_2}(\rb,\xb)=0$ for $\rb \in \Omega$ and for all $\xb$.
Since $M_{\rho_1-\rho_2}$ is analytic,
$M_{\rho_1-\rho_2}(\rb,\xb)=0$ for all $\xb$ and for $\rb \in \Omega$ implies that  $M_{\rho_1-\rho_2}(\rb,\xb)=0$ for all $\xb$ and all $\rb$, thus
$\tilde{M}_{\rho_1-\rho_2}(\nub,\mub)=0$ for all $\nub$ and $\mub$.
From Eq.~(\ref{FTM}), the nullity of $\tilde{M}_{\rho_1-\rho_2}$ is obtained only if 
$ \tilde{\rho_1}(\etab) - \tilde{\rho_2}(\etab) =0$ for $\etab \in W_{h^2}$.
Hence, if $\rho_1$ and $\rho_2$ have the same RIM variance, they have
the same spatial frequencies in $W_{h^2}$. Conversely, if two samples $\rho_1$ and $\rho_2$
 have the same frequency content in $W_{h^2}$, they generate the same variance image. 
We have thus demonstrated that
there is a one-to-one correspondence between the spatial frequencies of the variance
of diffraction-limited speckled images and the spatial frequencies of
the sample fluorescence density in the super-resolved Fourier domain  $W_{h^2}$
provided {\it the speckle autocovariance function $C$ be similar to the
  observation point spread function $h$}.

{\color{black} Now, what happens if the speckle autocovariance is different from the point spread function?}
If the speckle correlation length is larger than the width of the point spread function, {\it i.e.}
$W_E \subset W_h$, it is always possible to filter the recorded images so that
$h=C$. In this case, the variance of the modified images gives access to the sample spatial frequencies in $W_{E^2} $ at least.

On the other hand, if the speckle correlation length
is smaller than the width of
the point spread function, $W_h \subset W_E$, 
we show in appendix~\ref{B}, that, while the variance is sensitive to
the sample frequencies in $W_{hE}$,
it does not necessarily allow their recovery, even on the restricted domain $W_{h^2}$.
{\color{black} This remark applies in particular to configurations where the
  observation point spread function and the fluorescence density vary slowly over the speckle grain size.
This is the case in fluctuation imaging techniques
using near-field speckles and far-field detection \cite{choi2022} or optical speckles
and acoustic detection \cite{chaigne2016}.
In these techniques, hereafter called speckle-SOFI, the expression of the variance simplifies to,
\bal
 V_{\rm speckle-SOFI}(\rb) \approx C_0\int  h^2(\rb-\rb') \rho^2(\rb') \d\rb'.
 \label{SOFIRIM}
 \eal
 where $C_0=\int C(\rb)\rm{d}\rb $.}
 We observe that the variance is now linearly linked to the \textit{square} of the
sought parameter (optical absorption or fluorescence density) that is filtered over $W_{h^2}$.
Now, knowing the Fourier transform of $\rho^2$
in  $W_{h^2}$ does not mean that $\tilde{\rho}$ can be retrieved over $W_{h^2}$.
 Actually, when the
 speckled illumination is assumed to be spatially uncorrelated,
 it is possible to find samples  with different fluorescence density
 spectra in  $W_{h^2}$ that have the same variance image,  see appendix~\ref{C}.

 {\color{black} At this point, it is interesting to differentiate fluctuation
 imaging using quasi-uncorrelated speckled illuminations from
 Super-resolution Optical Fluctuation Imaging (SOFI). In SOFI, the intensity fluctuations
 observed in the recorded images come
 from the random activation of the fluorophores and not from the illumination
 (which is kept homogeneous and equal to $E_0$
 during the whole experiment).
 To account for this phenomenon, one needs to explicit
 further the characteristics of the fluorescence density $\rho$ which is
 related to the fluorophore concentration and the molecular brightness.
 We define the fluorophore concentration $g$ at the macroscopic scale such that,  $V g(\rb)$ is the number of fluorophores
 contained in a
macroscopic volume $V$ centered about $\rb$.
 Next, we introduce the mean molecular brightness $b$
 which accounts for the fluorophores' quantum yield and
for the environment-dependent ability of
the incident/emitted photons to reach the fluorophore/detector. If
all the fluorophores are activated in $V$, the mean brightness $b$ is defined such that
$ V g(\rb)b(\rb) E_0 $ is the energy measured by the camera
of the photons emitted from $V$. In other terms, if all the fluorophores are activated, the
fluorescence density is the product of the fluorophore concentration times the mean brightness, $\rho=g \times b$.

In SOFI, only a few fluorophores of $V$ are activated
during the image recording and they change at each novel image.
Let us assume that they follow a Poisson point process of intensity proportional
to the total number of fluorophores in $V$.
 Then, the number of activated
 fluorophores in $V$ observed when recording one image becomes
 a Poisson variable of parameter $V g(\rb)p(\rb)$ where
 $p$ is the mean percentage of activation.
 Under this assumption,
 we show in appendix~\ref{D} that the variance of SOFI images
 reads,
 \be
 V_{\rm SOFI}(\rb) =E_0^2\int h^2(\rb-\rb')  b^2(\rb') g(\rb')p(\rb')\d\rb'.
 \label{SOFI}
 \ee
 While RIM is able to recover the fluorescence density
 $\rho=g\times b$ over $W_{h^2}$,
 SOFI has a similar super-resolution capacity, but the latter applies to a
 distinct density $g \times b^2\times p=\rho\times b\times p$.}

{\color{black}  It is worth noting that, if the mean brightness $b$
  is homogeneous, RIM and
 SOFI are able to restore the fluorophore concentration
 $g$ over $W_{h^2}$, (provided the mean activation percentage
 $p$ in SOFI is also homogeneous).
 On the contrary, even if $b$ is homogeneous,
 fluctuation imaging using quasi-uncorrelated speckled
 illuminations (speckle-SOFI)
 can only restore the square of the fluorophore concentration,
 $g^2$, over $W_{h^2}$. Thus, SOFI and speckle-SOFI yield {\it a priori}
 different results and
their umbrella denomination as 'fluctuation imaging' can be misleading.}

In conclusion, we have shown that
the variance of images obtained under random speckled illuminations gives the same
fluorescence density as an ideal confocal microscope, {\it provided the speckle correlation length matches the width of the observation point spread function} (which is the case for RIM).
Our demonstration provides
 a solid theoretical ground to the
 two-fold resolution gain, the optical sectioning and the linearity to fluorescence observed in RIM \cite{Mangeat2021}.
 Also, we have pointed out the difference between
 the variance of SOFI-images obtained
 using the random activation of fluorophores and the variance
 of images obtained with quasi-uncorrelated speckled
 illuminations (speckle-SOFI).
 If the fluorophores brightness is homogeneous,
 SOFI variance is linearly linked to the fluorophore concentration
 while speckle-SOFI variance is quadratically linked to the fluorophore concentration.


%
%
%
%
\appendix
\section{ Proof of Identity }
\label{A}

Introducing Eqs.~(\ref{Buv},\ref{M}) in the left-hand term of Eq.~(\ref{norm}) yields,
\balx
A&=\int B_{U,V}(\rb) V(\rb) \d\rb
\ealx
where
\balx
&B_{U,V}(\rb)=\int  \d\rb_1 \d\rb_2 \d\xb
\\
&
h(\rb- \rb_1) U(\rb_1) h_E( \rb_1 - \xb)
h(\rb- \rb_2)
V(\rb_2) h_E( \rb_2 - \xb).
\ealx
Integrating over $\xb$ and using
$$
\int h_E(\rb_1 -\xb)h_E(\rb_2 -\xb)\d \xb= h(\rb_1-\rb_2)
$$
(where we have assumed that $C=h$) yields,
\balx
&B_{U,V}(\rb)=
\\&\int  \d\rb_1  \d\rb_2
h(\rb- \rb_1) U(\rb_1) h( \rb_1 - \rb_2) h(\rb- \rb_2) V(\rb_2).
\ealx
Then, decomposing $  h(\rb- \rb_2) = \int h_E(\rb -\xb) h_E(\rb_2 -\xb) \d\xb$, one obtains
\bal
\nonumber
&B_{U,V}(\rb)= \int  \d\rb_1 \d\rb_2 \d\xb
\\
\nonumber
&h(\rb  - \rb_1)   h_E( \rb - \xb)h(\rb_1- \rb_2) V(\rb_2) h_E( \rb_2 - \xb)U(\rb_1).
\eal
Finally, using the expression of $M_V$ in Eq.~(\ref{M}) and the symmetry of $h$ and $h_E$, we get
\bal
\nonumber
A= \int \left| M_V(\rb_1,\xb)\right|^2 U(\rb_1) \d\xb \d\rb_1.
\eal

\section{What happens when $W_h \subset W_E$ ?}
\label{B}
We have shown that provided $h=C$ (which can be obtained with an appropriate filtering of the raw images
as soon as $W_E \subset W_h$), there is a one to one correspondance between the
spatial frequencies of the variance of the speckled diffraction-limited images and that of the sample in $W_{h^2}$.
We now study a case where $h \neq C$ and  $W_h \subset W_E$.
To simplify the discussion, we assume that $(W_h, W_E)$ are centered
plain disks with frequency cut-offs $\nu_h$ and $\nu_E$ respectively, with $\nu_h<\nu_E$.
We further assume that the point spread function $h$
is symmetric so that  $\widetilde{h}$ is a real
positive symmetric function, like $\tilde{C}$.
We consider a sample whose Fourier spectrum is restricted to the null frequency and a high frequency $\pm \ve{k}$,
$\rho(\rb) = A + B\cos(2\pi \ve{k}\cdot \rb + \varphi )$ with $(A,B)$  real positive such that $\rho$ is real positive.
The variance of the raw images, given by Eq.~(\ref{var}), obtained with such sample reads,
\bal
\nonumber
V_{\rho}^{\rm RIM} (\rb) = &A^2\alpha + B^2\beta(\ve{k})+2AB \gamma(\ve{k})\cos(2\pi \ve{k}\cdot \rb +\varphi )
\\
&+B^2 \eta(\ve{k})\cos(4\pi\ve{k}\cdot\rb +2\varphi),
\eal
with
\bal
\nonumber
\alpha&=\int |\tilde{h}|^2(\nub)\tilde{C}(\nub)\rm d \nub,
\\
\nonumber
\beta(\ve{k})&=\int |\tilde{h}|^2(\nub+\ve{k})\tilde{C}(\nub)\rm d \nub,
\\
\nonumber
\gamma(\ve{k})&=\int \tilde{h}(\ve{k}-\nub)\tilde{h}(\nub)\tilde{C}(\nub)\rm d \nub,
\\
\nonumber
\eta(\ve{k})&= \int \tilde{h}(\nub)\tilde{h}(2\ve{k}- \nub) \tilde{C}(\ve{k}-\nub)\rm d \nub.
\eal
We observe that, as long as $k<\nu_h+\nu_E$, $\beta(\ve{k})\neq 0$ and the variance
depends on the high spatial frequency of the sample, $B$.
This result confirms the sensitivity of the variance
to sample spatial frequencies in $W_{hE}$.
However, if $2\nu_h<k \leq \nu_h+\nu_E$, $\gamma(\ve{k})=\eta(\ve{k})=0$
so that $V_{\rho}^{\rm RIM} (\rb)=\alpha A^2+\beta(\ve{k})B^2$.
In this case, the variance is sensitive to the amplitudes of the null and high frequencies of the sample, $(A,B)$,
but it has lost the information
about the phase of the high frequency, $\varphi$. 
 Worse, this
 exemple shows that a uniform sample defined by $\rho_1(\rb) =(A^2+\beta(\ve{k})B^2)^{\frac{1}{2}} /\alpha$
 will have the same variance as the
inhomogeneous sample defined by
$\rho(\rb) = A + B\cos(2\pi \ve{k}\cdot \rb + \varphi )$. Thus, when $\nu_h<\nu_E$, the
identifiability of the sample spatial frequencies from the variance is lost, even for frequencies belonging to $W_{h^2}$.
This assertion is particularly counter-intuitive as it shows that decreasing the size of the speckle grains
below the width of the observation point spread function is {\it a priori}
   detrimental to the sample reconstruction.

  {\color{black}  \section{Imaging with spatially quasi-uncorrelated speckles, speckle-SOFI}}
   \label{C}
   When the point spread function and fluorescence density are slowly varying over the
   width of the speckle autocovariance function, the variance of the diffraction-limited images
   is linearly linked to the square of $\rho$ filtered over $W_{h^2}$.
   In this section, we provide an example of two positive functions with  different
 Fourier contents in the super-resolved domain $W_{h^2}$ which, when squared,
 have exactly the same Fourier content in $W_{h^2}$. 

 We consider $g$ the sum of a constant and a one-dimensional cosine along the x-axis
 with a frequency $k$ laying in $W_{h^2}$ but not in $W_h$, and $f$ the sum of a constant and
 two cosines with period $k$ and $2k$. Note that $2k$ lays outside $W_{h^2}$.
 We adapt the constant and the cosine amplitudes so that $f$ and $g$ are positive
 and $f^2$ and $g^2$ are equal in
$W_{h^2}$.
  A possible solution is,
  \bal
  \nonumber
f(x) = 6 + \sqrt{2} \cos(2 \pi k x) + \cos(4 \pi k x) \\
g(x) = \frac{\sqrt{101} + 7}{2 \sqrt{2}} + \frac{\sqrt{101} - 7}{2} \cos(2 \pi k x)
\label{eq:ex_SOFI}
\eal
Noting $\mathcal{F}$ the low pass filter operator that removes
all the frequencies outside $W_{h^2}$, we find,
\bal
\nonumber
\mathcal{F}[f^2](x) = \mathcal{F}[g^2](x) = 75 + 13 \sqrt{2} \cos(2 \pi k x).
\eal

{\color{black} \section{Modeling SOFI at the macroscopic scale}
\label{D}

Generally,
SOFI data are modeled with a discrete sum that depends on
the fluorophore positions \cite{Dertinger2009}.
However, it is clearly impossible to recover the
fluorophores positions from SOFI (second-order) image,
except if a constraint of sparsity is assumed. It may thus be interesting to relate
 SOFI images to sample characteristics that are defined at a
macroscopic scale, such as the fluorophore concentration. 

In structured illumination microscopy,
one assumes that all the fluorophores are activated.
The fluorescence density $\rho$ is written as the product
of the fluorophore concentration $g$ with a mean brightness $b$. The
intensity recorded by the camera is modeled as,
\be
I(\rb)=\int h(\rb-\rb')E(\rb')g(\rb')b(\rb') \rm{d}\rb,
\ee
where $E$ is the inhomogeneous illumination intensity
and $h$ the microscope point spread function.

In SOFI, the illumination $E_0$ is homogeneous but the fluorophores
oscillate between an activated and non-activated state. Thus, only a subset of the
fluorophores present in the sample contributes to
the image intensity at a given time $t$.
The activated fluorophores in the (macroscopic) volume $V$  centered about $\rb$
can be seen as points popping up at random and
independently of one another. This
process is conveniently modeled with a Poisson point process of
intensity proportional to the number of
fluorophores in $V$.
Under this assumption, the
number of activated fluorophores in $V$ at time $t$, written as $V q(\rb,t)$, where $q$ is the
activated fluorophore concentration, is
a Poisson variable of parameter $V g(\rb)p(\rb)$ with
$p$ the mean percentage of activation.
With this definition, the intensity of the image recorded at $t$ can be written as,
\be
I(\rb,t)= E_0\int h(\rb-\rb') q(\rb',t)b(\rb') {\rm d}\rb'.
\label{sofi_model}
\ee
It is thus a filtered Poisson variable whose time variance reads \cite[Chap.\,5]{Snyder91},
\balx
V_{\rm SOFI}(\rb)&= E_0^2 \int h^2(\rb-\rb') b^2(\rb')g(\rb')p(\rb') {\rm d}\rb'.
\ealx
}

\bibliography{export}

\end{document}